\documentclass{elsart}

\usepackage{amssymb}
\usepackage{graphicx}
\usepackage{dcolumn}
\usepackage{bm}

\begin{document}

\begin{frontmatter}



\title{PI-CONTROLLED BIOREACTOR AS A GENERALIZED LI\'ENARD SYSTEM}


\author{V. Ibarra-Junquera and H.C. Rosu}

\address{Potosinian Institute of Science and Technology {\em (IPICyT)},\\
Apdo Postal 3-74 Tangamanga, 78231 San Luis Potos\'{\i}, Mexico}
\ead{vrani@ipicyt.edu.mx, hcr@ipicyt.edu.mx}

\begin{abstract}
\noindent It is shown that periodic orbits can emerge in Cholette's
bioreactor model working under the influence of a PI-controller. We
find a diffeomorphic coordinate transformation that turns this
controlled enzymatic reaction system into a  generalized Li\'{e}nard
form. Furthermore, we give sufficient conditions for the existence
and uniqueness of limit cycles in the new coordinates. We also
perform numerical simulations illustrating the possibility of the
existence of a local center (period annulus). A result with possible
practical applications is that the oscillation frequency is a
function of the integral control gain parameter.
\end{abstract}

\end{frontmatter}

\section{Introduction} \label{intro}

Mixing, understood as interpenetration of particles in different
zones of a given volume, is an important natural as well as
technological process. This is even more so when biochemical
reactions get involved. For the case of continuous stirred tank
reactors (CSTRs), Lo and Cholette \cite{Lo & Cholette} developed a
nonideal isothermal mixing model using a Haldane type chemical
reaction rate (which is similar to the Monod function for low
concentrations but includes the inhibitory effect at high
concentrations). This model has been studied extensively later by
many authors (\cite{Liou 1991 (Cholette Model)}, \cite{Padma 2002},
\cite{Chidambaram PI-Unstable Bioreactor 2003}, \cite{Chidambaram
Tuning PI-Controller Bioreactor 2003}, \cite{Kumar r=Kc/(1+KC)^2
(1994)}). In particular, Sree and Chidambaram \cite{Chidambaram
PI-Unstable Bioreactor 2003}, {\cite{Chidambaram Tuning
PI-Controller Bioreactor 2003} focused on the control problem by
means of a proportional integral (PI) control for this case. Indeed,
the PI controller is broadly used in the chemical and biochemical
industry. Therefore its closed-loop behavior is of much interest. In
this paper, we present a novel mathematical feature of this
closed-loop enzymatic reaction system, namely the possibility to be
represented as a dynamical system corresponding to a non polynomial
Li\'{e}nard oscillator. That means that given a PI-controlled CSTR
governed by the usual two-dimensional smooth dynamical system
\begin{eqnarray}
 \dot{X}=P(X,Y), \ \ \ \  \dot{Y}=Q(X,Y), \label{2D system}
\end{eqnarray}
we are able to find a diffeomorphic coordinate transformation
(Eq.~\ref{newcoord} below) that allows us to put it into the
well-known generalized Li\'enard form
\begin{eqnarray}
\dot{X} &=&  \phi(Y)-F(X)\label{Lienard1}\\
\dot{Y} &=& -g(X),\label{Lienard2}
\end{eqnarray}
where $g(X)$ is continuous on an open interval $(a_1,b_1)$, the
functions $F(X)$ and $\phi(Y)$ are continuously differentiable on
the open intervals $(a_1,b_1)$ and $(a_2,b_2)$, respectively. In
fact, these intervals can be extended to $-\infty \leq a_i < 0 < b_i
\leq \infty$, $i=1,2$.

In this paper, we show that the PI-controlled Cholette's CSTR model
belongs to this class of generalized Li\'{e}nard systems. Once doing
this, we make use of the beautiful results encountered in this
research area to study the periodic solutions near stationary points
for this particular application. Besides, the Hopf bifurcation is an
efficient way to study the existence of periodic orbits. In this
case, a pair of complex eigenvalues is assumed to exist and to cross
transversally the imaginary axis. Nevertheless, the fact that a Hopf
bifurcation guarantees the existence of a limit cycle does not imply
its uniqueness \cite{GAIKO 2000}. It is here where the uniqueness
result for Li\'{e}nard systems comes into play.

Thus, we extend the class of Li\'{e}nard-type system to the
interesting case of PI-controlled bioreactors for which the results
on the existence of limit cycles and their number as a function of
the control variables could be exploited in industrial applications.
In general, the study of oscillatory behavior in bioreactors is a
very important issue since it is generated by the coupled dynamics
of the most popular controller in industry (the PI one) and the
kinetics of the biochemical reactions. In addition, we shed light
here on an explicit example of a closed-loop system which is of
Li\'enard-type. Since the most direct way to interact with the
PI-controlled Cholette's CSTR is through the gain of the controller
the present analysis provide the users with definite conditions for
inducing oscillatory behaviors, which is instructive from the
pedagogical standpoint as well.

The paper is organized as follows. In Section 2, we discuss the
PI-controlled Cholette's bioreactor model and its basic assumptions.
In Section 3, we present the coordinate transformation that leads to
the Li\'{e}nard representation of this type bioreactor. The
existence of limit cycles is discussed in Section 4, and the
uniqueness consideration are included in Section 5. The numerical
simulation that we performed indicating the presence of the period
annulus are shortly described in Section 6. Finally, we end up the
work with several concluding remarks.

\section{Cholette's Dynamical Model} \label{Model}

The dynamical behavior without control actions (i.e., open-loop
operation) is governed by a unique nonlinear ordinary differential
equation (see Eq.~\ref{Modelo Cholette}). The non ideal mixing can
by described by the Cholette model \cite{Liou 1991 (Cholette
Model)}. This model was studied by Chidambaram \cite{Chidambaram
PI-Unstable Bioreactor 2003}, who proposed a tuning method for a
PI-controller. Examples, where this kind of kinetics occurs, can be
found in \cite{Kumar r=Kc/(1+KC)^2 (1994)}. The reactor model is
given by the following equations

\begin{eqnarray}
\frac {\mathrm{d}\zeta}{\mathrm{d}t} &=& \left(\zeta_f-\zeta \right)
\left[\frac {n F}{m V}\right]-\frac {K_1 \zeta}{\left(1+K_2 \zeta
\right)^2}~,\label{Modelo Cholette}
\end{eqnarray}
where the meaning of the parameters are given in table \ref{tab:1}.

\begin{table}[h]
 \centering \caption{Variables and parameters of Cholette's model.}
\label{tab:1}
\begin{tabular}{lll}
\noalign{\bigskip}\hline\hline\noalign{\smallskip}

    Symbol & \ \ \ \ \ \ \ \ \ \ \ Meaning & Units  \\

\noalign{\smallskip}\hline\noalign{\smallskip}

    $X$     & Substrate concentration       &\  $[ Kmol/m^3 ]$  \\
    $Y$     & Integrated error                &\  $[ Kmol\, s/m^3 ]$\\
    $F$     & Feed flow rate                &\  $[m^3/s]$                  \\
    $V$     & Volume                        &\  $[m^3]$                    \\
    $S_F$   & Substrate feed concentration  &\  $[ Kmol/m^3]$  \\
    $K_1$   & Maximal kinetic rate          &\  $[1/s]$                    \\
    $K_2$   & Inhibition parameter          &\  $[m^3/Kmol]$               \\
    $n$     & Mixing parameter              &\  $[$\footnotesize{\it dimensionless\,}$]$                  \\
    $m$     & Mixing parameter              &\  $[$\footnotesize{\it dimensionless\,}$]$                  \\
    $Kc$    & Proportional gain controller  &\  $[$\footnotesize{\it dimensionless\,}$ ]$                  \\
    $K_i$    & Integral gain controller      &\  $[$\footnotesize{\it dimensionless\,}$ ]$                  \\
    $u$     & Control input                 &\  $[$\footnotesize{\it dimensionless\,}$]$                  \\

\noalign{\smallskip}\hline
\end{tabular}
\end{table}
The following assumptions hold in Eq.~(\ref{Modelo Cholette}): (i)
all model parameters and physicochemical properties are constant
(ii) the reaction occurs in an nonideal mixed CSTR, operated under
isothermal conditions. The fraction of the reactant feed that enters
the region of perfect mixing is denoted by $n$, whereas $m$ denotes
the fraction of the total volume of the reactor where perfect mixing
is achieved. For $m$ and $n$ both equal to 1, the system is ideally
mixed. The values of the parameters $m$ and $n$ can be obtained from
the residence time distribution \cite{Liou 1991 (Cholette Model)}.
Fig.~\ref{Reactor} shows the schematic diagram of the bioreactor
configuration modeled by Eq.~(\ref{Modelo Cholette}).


\begin{figure}[h]
    \centering
    \includegraphics[height=6cm]{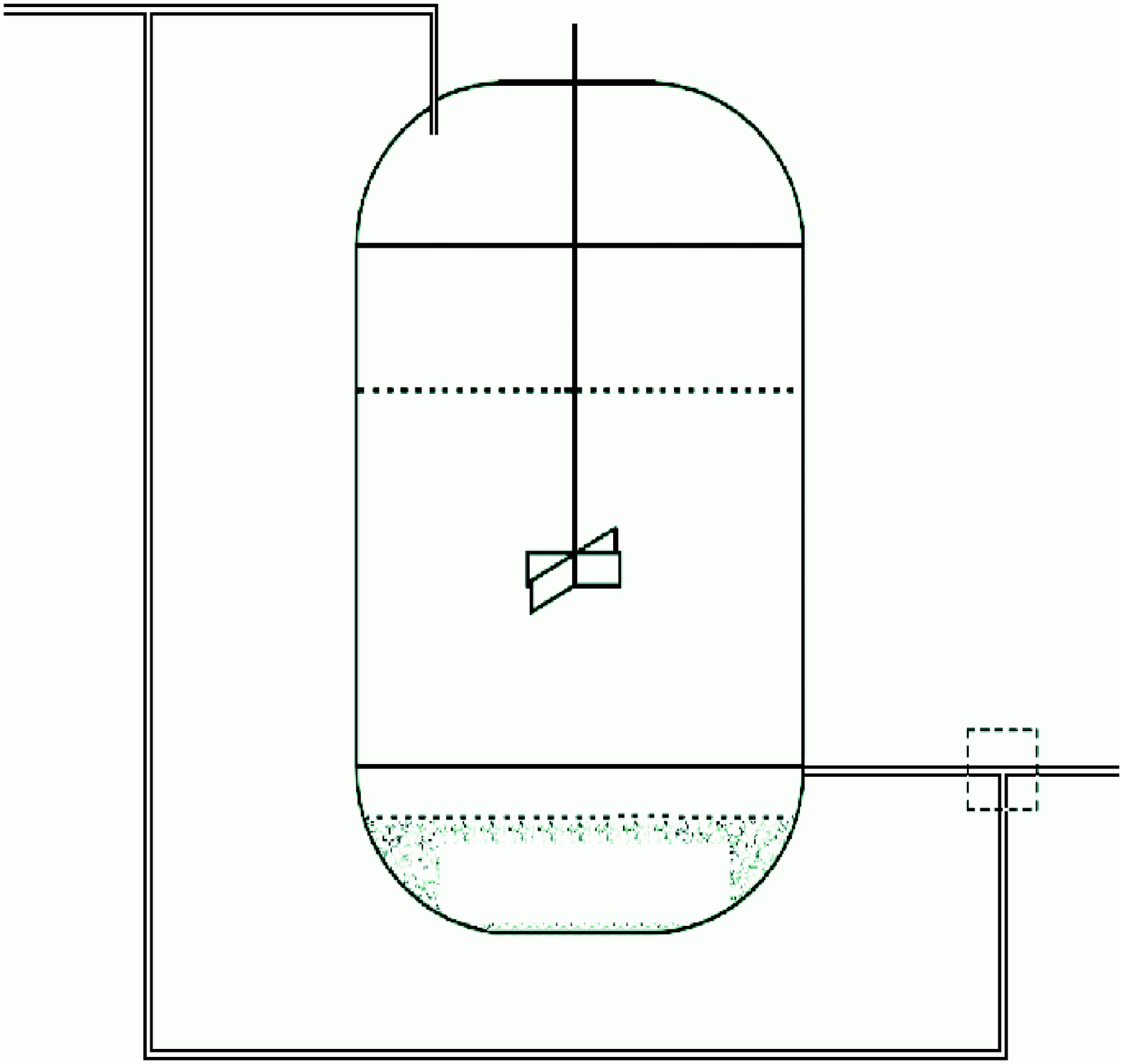}
        \put(-102,34){$(1-m)V$}
        \put(-145,163){$\zeta_f$}
        \put(-145,146){$nF$}
        \put(-180,163){$\zeta_f$}
        \put(-180,146){$F$}
        \put(-115,160){\line(-1,0){15}} \put(-115,160){\vector(1,0){1}}
        \put(-140,18){$(1-n)F$}
        \put(-130,3){$\zeta_f$}
        \put(-113,90){$\zeta$}
        \put(-113,80){$mV$}
        \put(15,52){\line(-1,0){15}} \put(15,52){\vector(1,0){1}}
        \put(-9,55){$\zeta'$}
        \put(-9,43){$F$}
        \put(-42,55){$\zeta$}
        \put(-42,43){$nF$}
        \put(-155,100){\line(0,1){15}} \put(-155,100){\vector(0,-1){1}}
        \put(-18,60){\line(0,1){15}} \put(-18,60){\vector(0,-1){1}}
        \put(-20,77){$\mathcal{A}$}
        \put(-85,18){\line(-1,0){15}} \put(-85,18){\vector(1,0){1}}
    \caption{Schematics of a classical continuous stirred tank bioreactor
with imperfect mixing corresponding to Cholette's model.}
\label{Reactor}
\end{figure}


Following the previous authors, we consider $\zeta_f$ as the
manipulated variable (i.e. $\zeta_f=\emph{u}$) and let $\zeta$ be
the controlled variable \cite{Chidambaram PI-Unstable Bioreactor
2003}. We are especially interested in the induced oscillatory
behavior of the bioreactor. The common control law in this case is
of the proportional integral type, that requires a dynamical error
extension in order to build the closed-loop two dimensional system.
Thus, the control law is given by
\begin{eqnarray}
\emph{u} \triangleq \left(-K_c \cdot Error - K_i \int Error
\,dt\right )~, \nonumber
\end{eqnarray}
where $K_c$ and $K_i$ are the control gain values. For the sake of
simplicity, the closed-loop system is written as:

\begin{eqnarray}
\dot{X} &=&  -X \left ( C+ \frac {K_1}{(1+K_2 X)^2} \right )
+ C \left (-{K_c} \left (X -{Ref}\right )-{K_i} Y\right )\label{sys1}\\
\dot{Y} &=& X-Ref~, \label{sys2}
\end{eqnarray}
where
\begin{eqnarray}
C=\frac{n\ F}{m \ V}~, \ \ \ \ \ \ \ \ \ \ \ X=\zeta~. \nonumber
\end{eqnarray}

Eq.~(\ref{sys1}) describes the dynamical behavior of the
concentration, while Eq.~(\ref{sys2}) refers to the dynamical
behavior of the integrated error.


\section{Transformation to the Li\'{e}nard form} \label{Lienard-Trans}

In this section, we show that the system given by
Eqs.~(\ref{sys1}-\ref{sys2}) can be rewritten as a system of the
form (\ref{Lienard1})-(\ref{Lienard2}), i.e., in the Li\'{e}nard
generalized form. This is one of the main results of this work.

{\bf Proposition 1.} {\small Under the transformation
\begin{eqnarray}
\left [ \begin{array}{c} {x}\\{y}
\end{array} \right ] = \Psi(X,Y) =
\left [ \begin{array}{c} {X-X_p}\\{-Y+Y_p}
\end{array} \right ], \label{newcoord}
\end{eqnarray}
where
\begin{eqnarray}
Y_p &=& -{\frac {{\it Ref}\, \left( C+K_{{1}}+{\it Ref}\,CK_{{2}}
\left( K_{{2 }}{\it Ref}+2 \right)  \right) }{C{\it K_i}\, \left(
1+K_{{2}}{\it Ref}
 \right) ^{2}}} \nonumber\\
X_p &=& Ref,\nonumber
\end{eqnarray}
system given by Eq.~(\ref{sys1}-\ref{sys2}) can be written in the
generalized Li\'{e}nard form. With the following additional
properties
\begin{itemize}
     \item[{[}A1{]}] $g(0)=0$ and $xg(x)>0$ for $x\neq0$;
     \item[{[}A2{]}] $\phi(0)=0$ and $\phi'(x)>0$ for $a_2<y<b_2$;
     \item[{[}A3{]}] The curve $\phi(y)=F(x)$ is well defined for all
     $x \in (a_1,b_1)$.
\end{itemize}}

{\bf Proof}. {\small If we substitute $X=x+X_p$ and $Y=-y+Y_p$ in
Eqs.~(\ref{sys1}) and (\ref{sys2}), and we choose
\begin{eqnarray}
F(x) &=& x \left( C \left( 1+{\it K_c} \right) +K_{{1}}\frac{ \left(
1-{\frac {K_{{2}} {\it Ref}\, \left( 2+K_{{2}} \left( x+2\,{\it Ref}
\right) \right) }{
 \left( 1+K_{{2}}{\it Ref} \right) ^{2}}} \right) }{ \left( 1+K_{{2}}
 \left( x+{\it Ref} \right)  \right) ^{2}} \right) \label{F(x)}\\
\phi(y) &=& C{\it K_i}\,y \label{phi(x)}\\
g(x) &=& x ~, \label{g(x)}
\end{eqnarray}
we get the generalized Li\'{e}nard  form of the PI-controlled
Cholette system. The properties {[}A1{]}, {[}A2{]} and {[}A3{]} are
straightforwardly checked in Eqs.~(\ref{F(x)})-(\ref{g(x)})}.

For $\Psi(X,Y)$ to be a diffeomorphism on the region $\Omega$, it is
necessary and sufficient that the Jacobian $\mathrm{d}\Psi(X,Y)$ be
nonsingular on $\Omega$, and moreover that $\Psi(X,Y)$ is one to one
from $\Omega$ to $\Psi(\Omega)$. Since $\Psi(X,Y)$ is linear, it is
one to one and the determinant of the Jacobian matrix is constant,
then is nonsingular in the region $\Omega=
[-\infty,\infty]\times[-\infty,\infty]$.

In the literature, the properties {[}Ai{]} are
standard properties assumed for Li\'{e}nard systems \cite{Xiao and
Zhang 2003}. We point out that the huge existing literature on
Li\'{e}nard systems deals mainly with cases in which $F(x)$ is
polynomial \cite{Giacomini 1998 (Ploy)}, \cite{Giacomini 1997
number}, \cite{Giacomini 1997 Improving}, whereas we are in a case
in which $F(x)$ is a nonlinear rational function. Such cases are far
less studied and there are still many open problems.

\section{Existence of Limit Cycles} \label{Hopf Bifurcation}
We briefly recall some basic results of the theory of bifurcations
of vector fields. Roughly speaking, a bifurcation is a change in
equilibrium points, periodic orbits, or in their stability
properties, when varying a parameter known as bifurcation parameter.
The values of the parameter at which the changes occur are called
bifurcation points. A Hopf bifurcation is characterized by a pair of
complex conjugate eigenvalues crossing the imaginary axis. Now,
suppose that the dynamical system $\dot{X}=f(X,\mu)$ with $X \in
\mathbb{R}^n$ and $\mu \in \mathbb{R}$ has an equilibrium point at
$X_{eq}$, for some $\mu={\mu}^{H}$; that is $f\left(
{X}_{eq},{\mu}^{H}\right)=0$. Let $A(\mu) =\frac{\partial f \left(
{X}^{H},{\mu}\right)}{\partial X}$ be the Jacobian matrix of the
system at the equilibrium point.
Assume that $A\left( {\mu}^{H}\right)$ has as single pair of purely
imaginary eigenvalues $S\left( {\mu}^{H}\right)= \pm
\dot{\imath}\omega_H$ with $\omega_H>0$ and that these eigenvalues
are the only ones with the properties $\Re(S)=0$. If the following
condition is fulfilled $
 \left. \frac{\mathrm{d}\Re\left(S(\mu)\right)}{\mathrm{d}
\mu} \right | _{\mu= \mu^{H}}~   \neq 0 \label{transver}
$ the Hopf bifurcation theorem states that at least one limit cycle
is generated at $\left(X_{eq},\mu^H\right)$ (see \cite{Romeo Ortega
Hopf in a Motor}). The condition (\ref{transver}) is known as the
transversality hypothesis. Considering now the $n$th degree
characteristic polynomial
$\lambda(S)=S^{n} + a_1 S^{n-1} + a_2 S^{n-2} + \ldots + a_n $,
where all the real $a_i$ coefficients are positive allows the
construction of a Hurwitz matrix $\mathcal{H}_{n \times n}$
Then one has the basic result that the characteristic polynomial is
stable if and only if the leading principal minors of
$\mathcal{H}_{n \times n}$ are all positive \cite{Ogata (4th
Edition)}.

To search the Hopf bifurcation we have to calculate the equilibrium
points of the system (\ref{Lienard1})-(\ref{Lienard2}) making equal
to zero the right-hand-side of the equation, and taking as
bifurcation parameters the values of the PI-control gains. Then,
finding the solution with respect to the state vector $x$ we notice
that the closed-loop system has a unique equilibrium point located
at the origin.

{\bf Proposition 2}.
{\small If the parameter $K_c$ is such that
\begin{eqnarray}
K_c^H-2\,{\frac {\sqrt {C{\it K_i}}}{C}}<{K_c}< K_c^H
~,\nonumber
\end{eqnarray}
then the Li\'{e}nard system (\ref{Lienard1}) and (\ref{Lienard2})
with the functions given in Eqs.~(\ref{F(x)})-(\ref{g(x)}) has at
least one limit cycle. The upper limit $K_c^H$ is defined in
Eq.~(\ref{Parameter bifurcation value}).}

{\bf Proof}.
{\small If we use the Hurwitz criterion to guarantee that the unique
equilibrium point is unstable, we evaluate the Jacobian of the
system at the origin
\begin{eqnarray}
\mathcal{J}\left(0,K_c\right)=\left[ \begin {array}{cc} -C-{\frac
{K_{{1}}}{ \left( 1+K_{{2}}{\it Ref} \right) ^{2}}}+2\,{\frac {{\it
Ref}\,K_{{2}}K_{{1}}}{ \left( 1+K_ {{2}}{\it Ref} \right)
^{3}}}-C{\it K_c}&C{\it K_i}\\\noalign{\medskip}- 1&0\end {array}
\right]~. \label{Jacobiano de neta}
\end{eqnarray}
Then, from the determinant $IS-\mathcal{J}(0,K_c)$, we get the
characteristic polynomial $\lambda(S)={S}^{2}+{a_1}{S}+{a_2}$, where
\begin{eqnarray}
a_{1}&=&C \left( 1+{\it K_c} \right) -{\frac {K_{{1}} \left(
K_{{2}}{\it Ref}-1
 \right) }{ \left( 1+K_{{2}}{\it Ref} \right) ^{3}}}
 \label{H1 Hurwitz}\\
a_{2}&=&{C}{K_i} ~.\label{H2 Hurwitz}
\end{eqnarray}
Then, the Hurwitz matrix is given by
\begin{eqnarray}
\mathcal{H}=  \left [ \begin{array}{cc}
{a_{1}} &{0}\\
{1} & {a_{2}}
\end{array} \right ]~, \label{Hurwitz del Systema}
\end{eqnarray}
and its principal minors are $\mathcal{H}_1 \left(K_c \right) =
{a_1}$ and  $\mathcal{H}_2 \left(K_c \right) = {a_1}{a_2}$. From
this formulas, we can note that the stability of the unique
equilibrium point depends on the sign of Eq.~(\ref{H1 Hurwitz}).
Since all the parameters in the Eqs.~(\ref{H1 Hurwitz}) and (\ref{H2
Hurwitz}) are positive, we can induce the local stability as a
function of the values of the controller gains involved in these
equations and then we obtain the bifurcation point as the trivial
solution of Eq.~(\ref{H1 Hurwitz}) for $K_c$ and restricting $K_i>0$
\begin{eqnarray}
{K_c}^H=-1+{\frac {K_{{1}} \left( K_{{2}}{\it Ref}-1 \right) }{C
\left( 1+K_{{ 2}}{\it Ref} \right) ^{3}}}~.
 \label{Parameter bifurcation value}
\end{eqnarray}
In order to test the transversality condition, the behavior of the
eigenvalues of $\mathcal{J}(0,K_c)$ in the neighborhood of $K_c^H$
should be analyzed. Thus, we take $K_c$ as $K_c^H+\epsilon$, with
$\epsilon \in {\mathbb{R}}$. The transversality condition will be
fulfilled if the sign of the equations $\mathcal{H}_1$ and
$\mathcal{H}_2$ changes when the sign of $\epsilon$ changes.
Substitution of Eq.~(\ref{Parameter bifurcation value}) in the
principal minor expressions gives $ \mathcal{H}_1\left(K_c^H\right)
= {\epsilon}\left({C}\right)$ and $\mathcal{H}_2\left(K_c^H\right) =
{\epsilon}\left({K_i}{C}^2\right)$.

 From the above equations we can
appreciate that if we want to have a positive real part of the
eigenvalues of the Jacobian matrix (\ref{Jacobiano de neta}), we
need that $\epsilon$ be negative. In other words, it is below the
value given by Eq.~(\ref{Parameter bifurcation value}) where the
limit cycles are generated. Using $K_c^H$, the eigenvalues of the
matrix (\ref{Jacobiano de neta}) are given by the roots of the
characteristic polynomial $\lambda (S)$, which are
\begin{eqnarray}
S_1 &=&  - \frac{\epsilon\,C}{2}+ \frac{\sqrt
{{\epsilon}^{2}{C}^{2}-4\,C{\it K_i}}}{2} \\
S_2 &=& - \frac{\epsilon\,C}{2}- \frac{\sqrt
{{\epsilon}^{2}{C}^{2}-4\,C{\it K_i}}}{2}~,
\end{eqnarray}
where we can notice that the eigenvalues are complex with positive
real parts, when $0>\epsilon>-2\,{\frac {\sqrt {C{\it K_i}}}{C}}$
and $K_i>0$.}

Eq.~(\ref{Parameter bifurcation value}) is of main importance,
because it corresponds to the Hopf bifurcation and therefore lies in
a neighborhood of the value of the parameter where at least one
limit cycle is generated. Note, that the Hopf bifurcation by itself
can not guarantee the uniqueness of the limit cycle, because more
than one limits cycle could appear \cite{GAIKO 2000}. Then we need
to use additional constraints in order to find the condition for
uniqueness.

\section{Uniqueness of Limit Cycles}
Xiao and Zhang \cite{Xiao and Zhang 2003} gave an interesting
theorem on the uniqueness of limit cycles for generalized
Li\'{e}nard systems, under the conditions {[}A1{]}, {[}A2{]} and
{[}A3{]} that allows us to prove a novel property of PI-controlled
Cholette bioreactors.

{\bf Theorem 1}. {\small Using the notations $G(x)=\int
_{0}^{x}g(x)dx$ and $f(x)=F^{'}(x)$, suppose that the system
(\ref{Lienard1}) - (\ref{Lienard2}) satisfies the following
conditions:

\begin{itemize}
     \item[{(}\textbf{i}{)}] there exist $x_1$ and $x_2$,
     $a_1<x_2<0<x_1<b_1$ such that $F(x_1)=F(0)=0$, $F(x_2)>0$ and
     $G(x_2)\leq G(x_1)$; $xF(x)\leq 0$ for $x_2\leq x \leq
     x_1$, $F'(x) > 0$ for $a_1 < x <x_2$ or $x_1 < x <b_1$, and
     $F(x) \neq 0$ for $0 < \mid x \mid \ll 1$.
     \item[{(}\textbf{ii}{)}] $F(x)f(x)/g(x)$ is nondecreasing for $x_1<x<b_1$.
     \item[{(}\textbf{iii}{)}] $\phi'(y)$ is nonincreasing as $\mid y \mid$
     increases.
\end{itemize}

Then the system (\ref{Lienard1}) and (\ref{Lienard2}) has at most
one limit cycle, and it is stable if it exists. }

Note that the above result does not guarantee the existence of limit
cycles by itself.

{\bf Proposition 3}. {\small System (\ref{Lienard1}) and
(\ref{Lienard2}) with the functions given by the Eqs.~
(\ref{F(x)})-(\ref{g(x)}) and with $Ref>{2}/{K_2}$ and
$K_c={K_c}^H+\epsilon^{\ast}$, where
\begin{eqnarray}
\epsilon^{\ast}=-\,{\frac {K_{{1}} \left( -2+K_{{2}}{\it Ref}
\right) }{2C \left( 1+ K_{{2}}{\it Ref} \right) ^{3}}}~,
\end{eqnarray}
has a unique limit cycle.}

{\bf Proof}. {\small The existence is given by the Proposition 2,
and the uniqueness will be proved by showing that the system
fulfills the conditions of Theorem 1.
Let us take $Ref^{\ast} ={2\kappa}/{K_2}$ with $\kappa > 1$ which
agrees with the condition stated in Proposition 3. Then
\begin{eqnarray}
{K_c}^H(\epsilon^{\ast},Ref^{\ast})=-1+{\frac {K_{{1}}\kappa}{C
\left( 1+2\,\kappa \right) ^{3}}}~. \nonumber
\end{eqnarray}
For these values of the parameters, the function $F(x)$ is
simplified to the form
\begin{eqnarray}
F^{\ast}(x)={\frac {xK_{{1}} \left(
3\,\kappa+\kappa\,{K_{{2}}}^{2}{x}^{2}-4\,{ \kappa}^{3}+1 \right) }{
\left( 1+2\,\kappa \right) ^{3} \left( 1+K_{{ 2}}x+2\,\kappa \right)
^{2}}}~. \label{F_esp(x)}
\end{eqnarray}
The real roots of $F^{\ast}(x)$ are $0$, $x_1$, and $-x_1$, where
\begin{equation}
x_{1} = {\frac {\sqrt {\kappa\, \left( -1+\kappa
\right) } \left( 1+2\,\kappa
 \right) }{\kappa\,K_{{2}}}}~,\nonumber
\end{equation}
which fulfill the property $F(x_1)=F(0)=0$. Moreover, taking
$x_2=-\xi x_1$, where $0<\xi<1$, one can see that $x_2<0<x_1$. To
evaluate $F^{\ast}(x)$ in the intervals $x_2<x<0$ and $0<x<x_1$, it
is sufficient to substitute in Eq.~(\ref{F_esp(x)}) $x=\beta \,
x_{i}$, where $i=1,2$ and $0<\beta<1$. Then we can easily verify
that

\begin{eqnarray}
0>F^{\ast}(\beta\, x_1) &=& {\frac { \left( -1+\kappa \right)
\left( \beta ^2-1 \right) \kappa\,K_{{1}}\beta\,\sqrt {\kappa\,
\left( -1+ \kappa \right) }}{ \left( 1+2\,\kappa \right) ^{2} \left(
\kappa+ \beta\,\sqrt {\kappa\, \left( -1+\kappa \right)
} \right) ^{2}K_{{2}} }}\\
0<F^{\ast}(\beta \,x_2) &=& -\xi {\frac { \left( -1+\kappa \right)
 \left( \beta ^2\,-1 \right) \kappa\,K_{{1}}\,\beta\,\sqrt {
\kappa\, \left( -1+\kappa \right) }}{ \left( 1+2\,\kappa \right)
^{2}
 \left( -\kappa+\,\beta\,\sqrt {\kappa\, \left( -1+\kappa \right)
} \right) ^{2}K_{{2}}}}~.
\end{eqnarray}
Consequently, $xF^{\ast}(x)\leq 0$ for $x_2\leq x \leq x_1$.
Moreover, $G(x)=x^2/2$ and then $G(x_2)\leq G(x_1)$. Besides, for
$\theta>1$ we have
\begin{eqnarray}
0<F^{\ast '}({\theta}{x_1})={\frac { \left( -1+\kappa \right)
\left((-1+3\,{\theta}^{2}) \kappa+(\theta + \theta ^3)\sqrt
{\kappa\, \left( -1+\kappa \right) }
\right) K_{{1}}{\kappa}^{2} }{ \left(
1+2\,\kappa \right) ^{3} \left( \kappa+\theta\sqrt {\kappa\,
 \left( -1+\kappa \right) }\right) ^{3}}}~. \nonumber
\end{eqnarray}
In other words, $F^{\ast '}(x) > 0$ for $x_1 < x <b_1$. From
Eq.~(\ref{F(x)}) we can see that $F^{\ast}(x) \neq 0$ for $0 < \mid
x \mid \ll 1$. In addition, since $\phi'(y)=C K_i$, no matter how
$\mid y \mid$ increases, $\phi'(y)$ remains constant. Finally, we
can write $\Psi(x)\equiv
F^{\ast}(x)f(x)/g(x)=(\mathcal{A}+\mathcal{B})/\mathcal{C}$, where
\begin{eqnarray}
\mathcal{A} &=& 2+\kappa\, \left(
16\,{\kappa}^{2}+{K_{{2}}}^{3}{x}^{3} \right) - \kappa\, \left(
1+2\,\kappa \right)  \left( xK_{{2}}
\left( 3\,K_{{2}} x+8\,\kappa+4 \right) -12 \right)\\
\mathcal{B} &=&  2\,\ln  \left( 1+K_{{2}}x+2\,\kappa \right)  \left(
1+2\,\kappa  \right) ^{3} \left( 1+K_{{2}}x+2\,\kappa \right) \\
\mathcal{C} &=& {\frac {{K_{{1}}}^{2}}{2 \left( 1+2\,\kappa \right)
^{6}{K_{{2}}}^ {2} \left( 1+K_{{2}}x+2\,\kappa \right) ^{3}}}~.
\end{eqnarray}
Performing the derivative of $\Psi(x)$, evaluating it at $x
={x_1}{\alpha}$ for $\alpha > 1$, and plotting $\vartheta
\Psi'({x_1}{\alpha})$, where
\begin{equation}
\vartheta={\frac { \left( 1+2\,\kappa \right) ^{5}
 \left( \kappa+\alpha\,\sqrt {\kappa\, \left( -1+\kappa \right) }
 \right) ^{4}K_{{2}}}{{K_{{1}}}^{2}{\kappa}^{3}}}~,
\end{equation}
we get the positive function displayed in Fig.~\ref{DF}. Thus,
$F^{\ast}(x)f(x)/g(x)$ is nondecreasing for $x_1<x<b_1$. In this
way, we
checked each of the three conditions of Theorem 1. 
The unique limit cycle can be seen in Fig.~\ref{unique} that
illustrates the numerical simulation corresponding to the results
obtained until now.
\section{Case Study}
With the aim of illustrate the results given in the previous
sections, we perform numerical simulation using the values of the
parameters given by Chidambaram \cite{Chidambaram Tuning
PI-Controller Bioreactor 2003}. All our calculations are performed
for a flow characterized by the value of the Damk\"ohler number
$(Da=K_1V/F)$ equal to 300, as reported by Sree and Chidambaram
\cite{Chidambaram PI-Unstable Bioreactor 2003}.

\begin{table}[h]
\centering \caption{The values of the parameters of Cholette's model
\cite{Chidambaram PI-Unstable Bioreactor 2003}.} \label{tab:2}
\begin{tabular}{lll}
\noalign{\bigskip}\hline\hline\noalign{\smallskip}

    Symbol & Value & Units  \\

\noalign{\smallskip}\hline\noalign{\smallskip}

    $F$     &\  $3.333\times 10^{-5}$  &\  $[m^3/s]$                  \\
    $V$     &\  $10^{-3}$        &\  $[m^3]$                    \\
    $K_1$   &\  $10$             &\  $[1/s]$                    \\
    $K_2$   &\  $10$             &\  $[m^3/Kmol]$               \\
    $n$     &\  $0.75$       &\  $[$\footnotesize{\it dimensionless\,}$]$                  \\
    $m$     &\  $0.75$       &\  $[$\footnotesize{\it dimensionless\,}$]$                  \\

\noalign{\smallskip}\hline
\end{tabular}
\end{table}
In Fig.~\ref{frequency} a plot of the evolution in time of the
variable $x$ is given for two values of the integral parameter
$K_i$. It indicates that the oscillation frequency can be
manipulated through this control parameter.
\begin{figure}
    \centering
    \includegraphics[height=9cm]{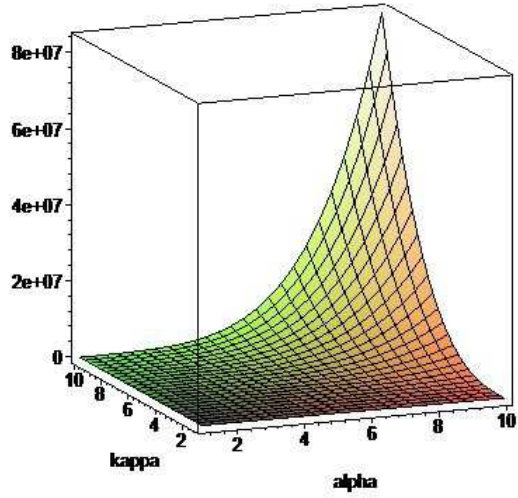}
    \caption{The function $\vartheta
\Psi'(\alpha x_1)$, for $\alpha > 1$, and $\kappa>1$. We see that
this is a strictly positive function.}
    \label{DF}
\end{figure}

}

\begin{figure}
    \centering
    \includegraphics[height=7cm]{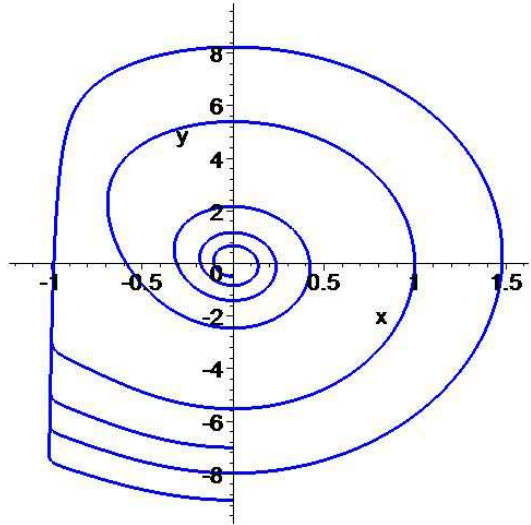}
    \caption{The phase portrait
     of the PI-controlled Cholette model subjected
      to the uniqueness conditions. The employed values
      of the parameters are those given in Table \ref{tab:1}.}
    \label{unique}
\end{figure}

\begin{figure}[h]
    \centering
    \includegraphics[height=8cm]{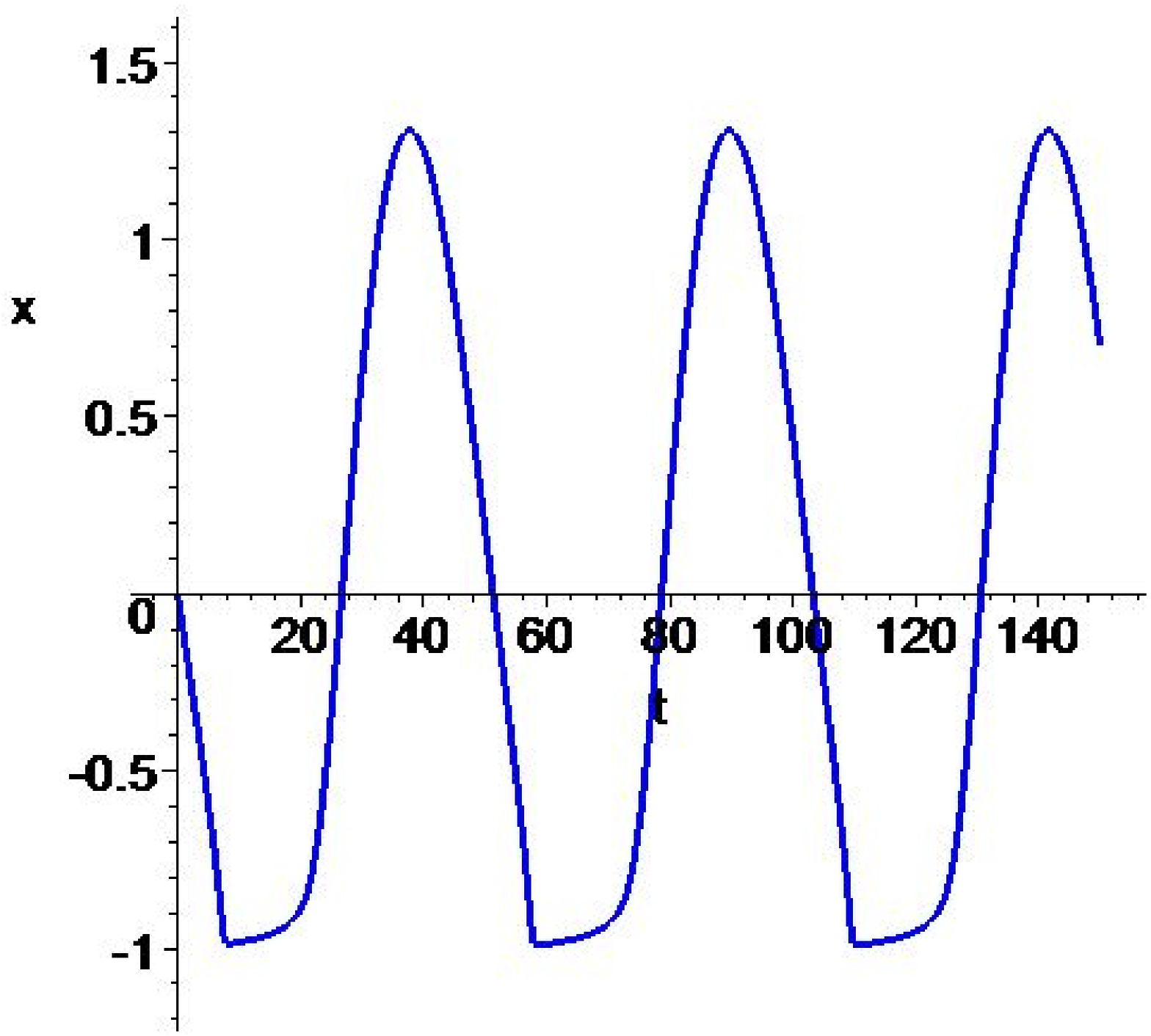}
    \includegraphics[height=8cm]{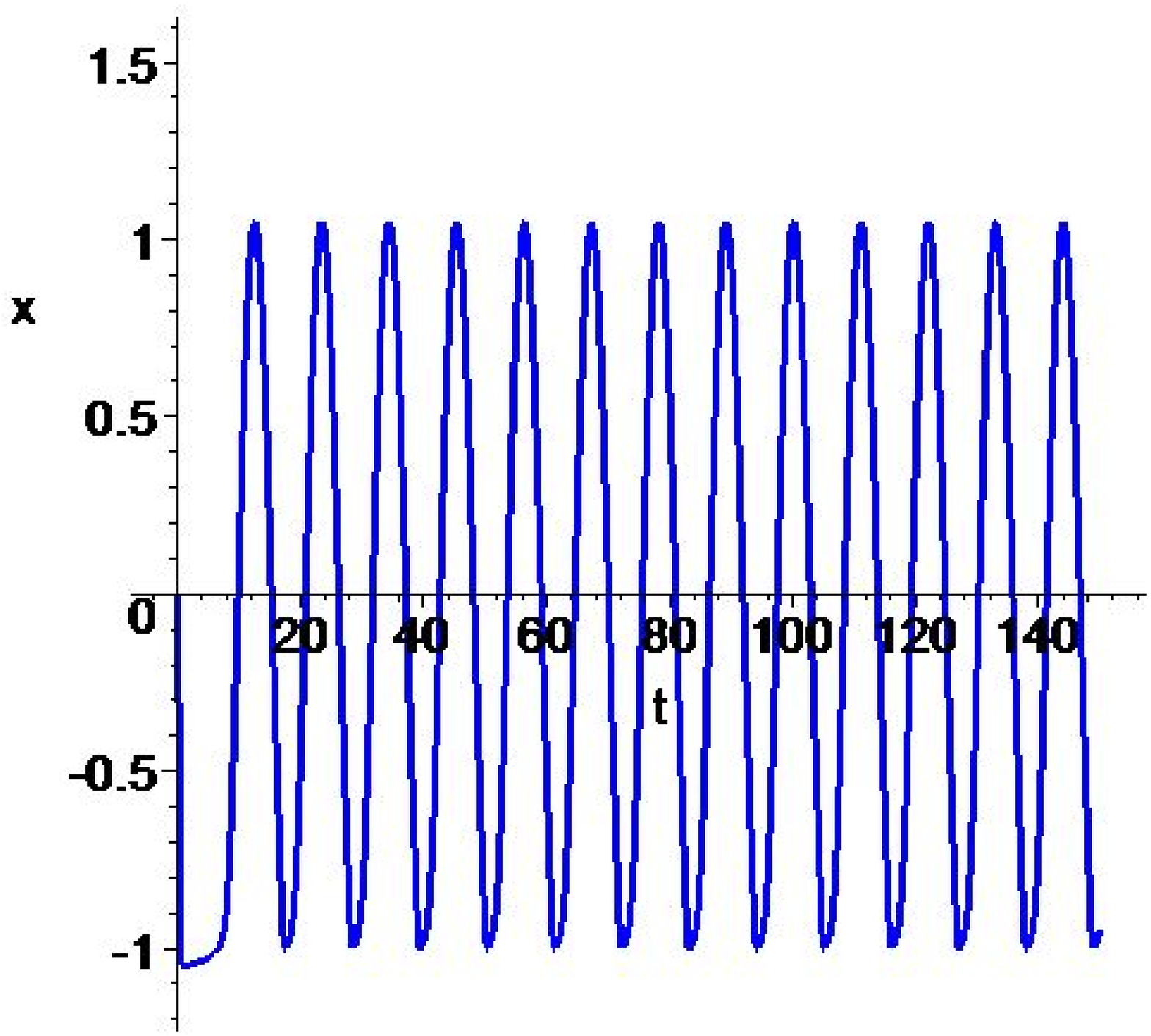}
    \caption{Time evolution of the variable $x$. The left hand
     side plot corresponds to $K_i=0.5$, while the right hand side plot corresponds to $K_i=10$.
      This is a graphical representation of the fact that oscillation frequency is a function of the control gain parameter $K_i$.}
    \label{frequency}
\end{figure}

\newpage
\section{A Local Center of the Generalized Li\'{e}nard System ?} \label{annulus}

We begin this section by recalling that a \emph{limit cycle} is an
isolated closed orbit, while a critical point is a \emph{center} if
all orbits in its neighborhood are closed. To the best of our
knowledge the literature on period annuli for Li\'{e}nard systems is
well developed only for polynomial cases and moreover it focuses on
Hamiltonian type systems \cite{Llibre 2004}, \cite{Du Z. 2004},
\cite{Cristopher 1997}.

Notice that for $K_c(Ref^{\ast})=K_c^H(Ref^{\ast})+\epsilon$ and
$\epsilon$ in the following interval $\epsilon ^{\ast}<\epsilon <0$
the existence of limit cycles is proved but without guaranteeing
uniqueness. It is precisely in this interval where our numerical
simulations point to the existence of a local center in a
neighborhood of the origin. Fig.~\ref{Comportamiento Oscilatorio}
shows the phase portrait of the PI-controlled generalized Li\'enard
system with a value of $K_c$ in the same interval and close to
$K_c^H$.

\begin{figure}[h]
    \centering
    \includegraphics[height=7.2cm]{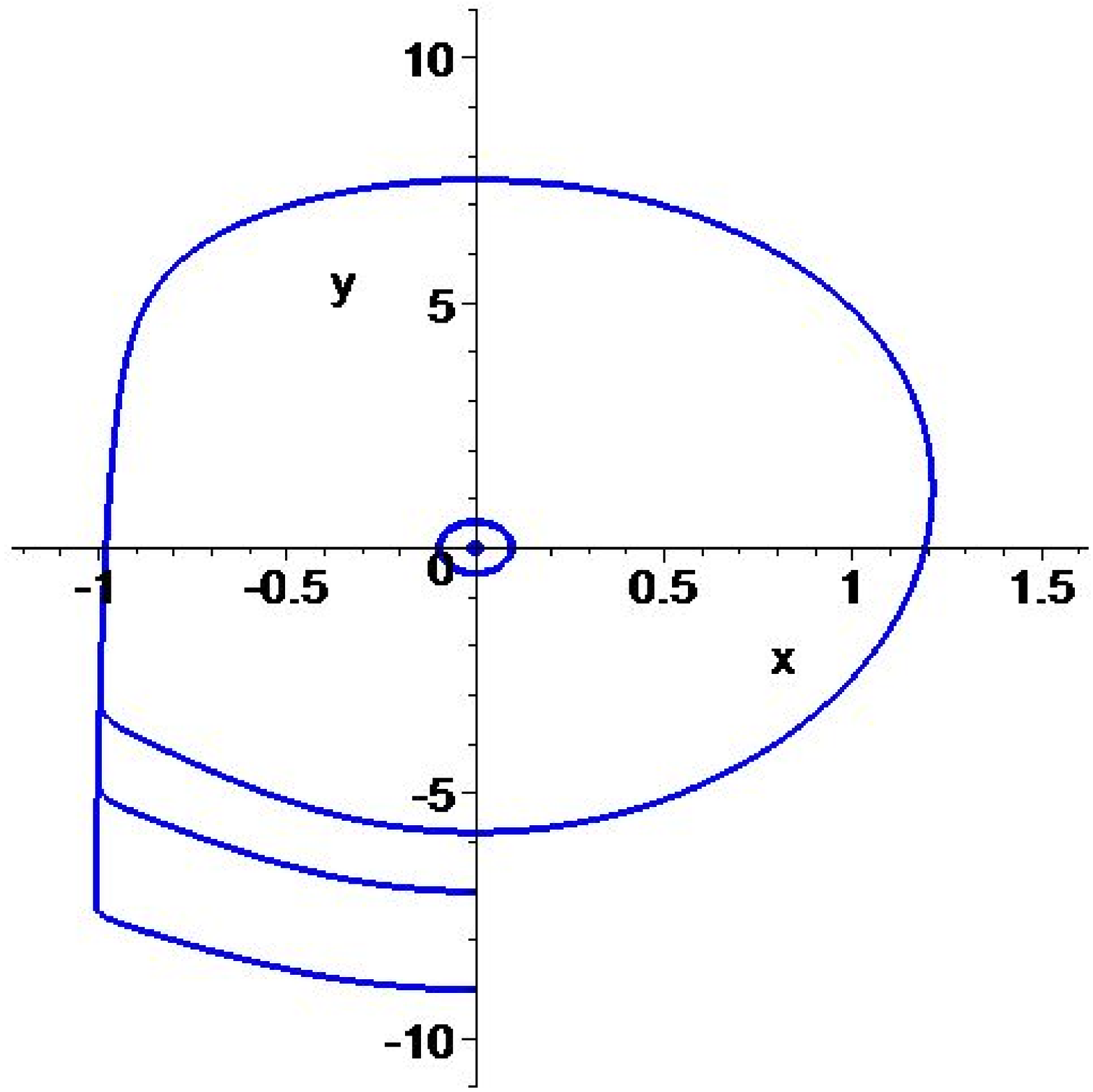}
    \includegraphics[height=7.2cm]{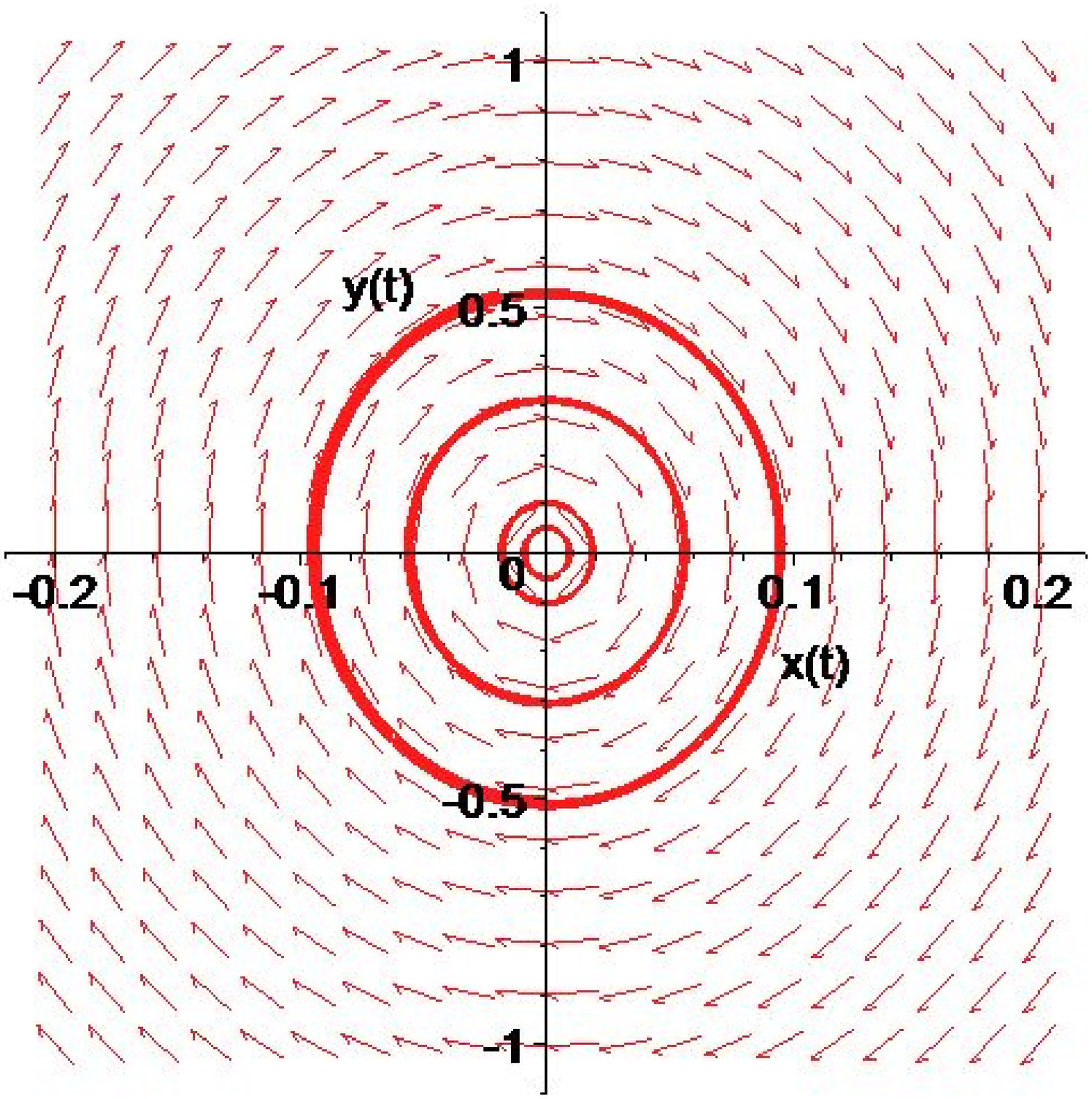}
    \caption{The phase portrait of the PI-controlled generalized Li\'enard
system with a value of $K_c^H(Ref^{\ast})+\epsilon$ in the interval
$\epsilon ^{\ast}<\epsilon <0$
and close to $K_c^H$. The upper plot shows
the state space with the limit cycle and the possible annulus in the
neighborhood of the origin, while the lower plot displays the
configuration of the vector field close to the origin.}
    \label{Comportamiento Oscilatorio}
\end{figure}

\section{Conclusions} \label{annulus}

The main result of this paper is that the Cholette CSTR model under
PI control can be mapped into a generalized Li\'enard dynamical
system of nonpolynomial type. Thus, we establish a new important
application of this class of nonlinear oscillators that allows us to
make a detailed study of the oscillatory dynamical behavior of these
interesting bioreactors.

Sufficient conditions for the existence and uniqueness of limit
cycles of this generalized Li\'enard system are stated in this paper
together with numerical simulations that indicate the possibility of
the existence of a local center (period annulus) when the gain
proportional parameter $K_c$ of the control law is close to the
value $K_c^H$ corresponding to the existence condition of limit
cycles. We also notice that the oscillation frequency is a function
of the integral control gain parameter $K_i$, a result that could
have practical applications. We mention that similar results have
been obtained by Albarakati, Lloyd, \& Pearson \cite{ejde00} for the
polynomial case.

Our work also shows that the Li\'enard representation of dynamical
systems and its associated results could have a remarkable potential
as an effective tool in the control theory for the closed-loop
dynamical analysis in the plane.

\section*{Acknowledgements}

This work has been partially supported by CONACyT project 46980.

\label{}



\end{document}